\documentclass[
]{ceurart}

\begin{document}

\copyrightclause{Copyright for this paper by its authors.
  Use permitted under Creative Commons License Attribution 4.0
  International (CC BY 4.0).}

\conference{First Large Language Models for Model-Driven Engineering Workshop (LLM4MDE 2024), Enschede, Netherlands}

\title{From Image to UML: First Results of Image-Based UML Diagram Generation using LLMs}


\author[1]{Aaron Conrardy}[%
orcid=0000-0002-3030-4529,
email=aaron.conrardy@list.lu
]
\cormark[1]
\address[1]{Luxembourg Institute of Science and Technology, Luxembourg}

\author[1,2]{Jordi Cabot}[%
orcid=0000-0003-2418-2489,
email=jordi.cabot@list.lu
]
\address[2]{University of Luxembourg, Luxembourg}

\cortext[1]{Corresponding author.}

\begin{abstract}
In software engineering processes, systems are first specified using a modeling language such as UML. These initial designs are often collaboratively created, many times in meetings where different domain experts use whiteboards, paper or other types of quick supports to create drawings and blueprints that then will need to be formalized. These proper, machine-readable, models are key to ensure models can be part of automated processes (e.g. input of a low-code generation pipeline, a model-based testing system, ...). But going from hand-drawn diagrams to actual models is a time-consuming process that sometimes ends up with such drawings just added as informal images to the software documentation, reducing their value a lot. To avoid this tedious task, we explore the usage of Large Language Models (LLM) to generate the formal representation of  (UML) models from a given drawing. More specifically, we have evaluated  the capabilities of different LLMs to convert images of (hand-drawn) UML class diagrams into the actual models represented in the images. While the results are good enough to use such an approach as part of a model-driven engineering pipeline we also highlight some of their current limitations and the need to keep the human in the loop to overcome those limitations. 
\end{abstract}


\maketitle

\section{Introduction}

The continuous progress and rise of Large Language Models (LLM) has led to their integration into various tasks in various domains, such as medical advice consultation in healthcare, writing or reading assistance in education, legal interpretation and reasoning in Law or financial reasoning \cite{zhao2023survey}. Computer science is not an exception, with various LLMs being used in software development as programming assistants \cite{barke2022grounded}, but also in software modeling as a model creation tool from natural language \cite{texttouml, texttouml2}, showing promising results on both fronts. 
Beyond textual input, some LLMs additionally provide support for image input, also described as visual LLMs. 
Generally, OpenAI's \textbf{GPT-4V}\footnote{\url{https://openai.com/research/gpt-4v-system-card}} and \textbf{Google's Gemini (Pro/Ultra)}\footnote{\url{https://deepmind.google/technologies/gemini}} are seen as the best publicly available multimodal LLM. Unfortunately, these are either hidden behind a paywall or only accessible through specific platforms' interfaces. \textbf{CogVLM} \cite{wang2023cogvlm} acts as an open-source alternative, providing not only its source code but also a free chat interface to interact with the model.
These multi-modal LLMs further broadened the range of possible applications, such as tools like Design2Code \cite{si2024design2code} or Make-real\footnote{\url{https://github.com/tldraw/make-real}} that allow for the generation of HTML, CSS and JS code from either screenshots of web pages or mock-ups. 
In the domain of software modeling, we believe that these image reading capabilities could also be leveraged to facilitate and accelerate the modeling process itself. 

Indeed, one of the biggest hurdles of modeling tools is the usability of such tools \cite{usabilitytool}. Take the example of UML class diagrams that are used to represent the structure of software.
These are usually collaboratively sketched on a whiteboard in an attempt to crystallize the initial design idea of an application. 
Yet, once drawn, there still is the need to transform the drawing into a more polished form, for example for documentation purposes. 
Additionally, one might also want a computer readable format to use the UML class diagram for code generation.
The same goes for migrating legacy projects where, most likely, no specification is available but rather only a few drawings as part of the system documentation. 

In a low-code context, which focuses on reducing the amount of hand-coding required to create applications, these models can be further processed to (semi)automatically generate software. Following these footsteps, the term "low-modeling" \cite{cabot2024lowmodeling} was also created to describe techniques and tools that accelerate the modeling process, effectively speeding up the low-code pipeline.
We believe that a tool capable of converting given images to UML diagrams could speed up this tedious process and quickly provide initial models that could be extended or further used, effectively enabling the low-modeling of software. 

In this paper, we explore the capabilities of LLMs with image processing capabilities to act as image to UML converters, contributing towards the hybridization between Software Engineering and LLMs \cite{fan2023large}.
We conducted experiments on various images of drawn UML diagrams using visual LLMs and evaluate the correctness and completeness of the generated results. 
Our findings reveal that GPT4-V provided the best results in terms of correctness and completeness, and indeed that the usage of LLMs to convert images to UML provides positive results, although with the need of keeping the human in the loop and dependent on the used LLM. 

The rest of the paper is structured as follows: in Section \ref{sota}, we go over related work and the used approaches. Section \ref{rq} lists the research questions. Section \ref{exp} describes the experiment, the results and interprets them. Section \ref{discussion} consists of a discussion concerning the results and further findings. Section \ref{toolsupp} describes a tool with an image to UML feature. Section \ref{final} concludes the paper and describes the planned future work.

\section{State of the art}\label{sota}

Existing research works have already tackled the concept of generating UML diagrams in a computer readable format from given images of diagrams. Most notably, \cite{image2umlocl} proposes the Img2UML tool that aimed at generating XMI files from images of UML class diagrams. The tool was specifically tailored to recognize UML class diagrams created with computer-aided software engineering (CASE) tools and makes use of optical character recognition (OCR) techniques for analysing the provided images. It is unclear whether the tool also works for drawn UML class diagrams and we could not access the tool to conduct our own tests. Another more recent attempt at recognizing images of UML class diagrams is ReSECDI \cite{ReSECDI}. Yet, ReSECDI only focuses on recognizing semantic elements, such as classes or relationships, from given images and generates an output text file they describe as semantic design model. The output contains the recognized semantic information, but does not follow a standard notation for UML class specification. Again, it is unclear whether ReSECDI could work with hand-drawn diagrams. 

Other works abstract the task further by either only providing a set of information from given diagrams \cite{umlrec2}, such as the location and text inside classes, or only attempt to classify the type of UML diagram and not the content itself using deep learning techniques \cite{classifyUML}. 
Not a lot of research focuses on the ability to extract information from images of hand-drawn UML diagrams and existing attempts are outdated and do not hold up to the constant technological progress or focus on drawings made with a specific tool and not on actual paper \cite{drawnuml2model1, drawnuml2model2}.

As previously mentioned, attempts at generating UML models using LLMs have already been made \cite{texttouml, texttouml2} using the PlantUML notation as output, yet, these two attempts only tackle textual input. To the best of our knowledge, we are the first to leverage LLMs' image recognition capabilities to generate UML models.

\section{Research questions}\label{rq}
Our main goal is to evaluate vision LLMs' capabilities to process and transform images of UML diagrams into a computer readable format while also exploring possible variables that affect the transformation. We focus on hand-drawn UML diagrams as we assume that the transformation of images of hand-drawn UML diagrams into a computer readable format is more difficult than the same task for images of UML diagrams that were created with CASE tools, as drawings generally contain more inconsistencies in terms of handwriting, lines, etc. This assumption implies that any results obtained with images of hand-drawn UML diagrams can be used as an approximate result for images of UML diagrams created with CASE tools. Moreover, we will primarily focus on UML class diagrams, as these seem to be the most popular type of UML diagram \cite{usedUML2}. 
We partially inspire ourselves from the experiment conducted in \cite{texttouml2} and formulate the following Research Questions (RQ):
\begin{itemize}
    \item \textbf{RQ1}: Are LLMs capable of providing a complete (classes, relationships, textual content,...) re-creation of a given image containing the depiction of a UML class diagram?
    \item \textbf{RQ2}: Do LLMs respect the syntax of the chosen notation for the output?
    \item \textbf{RQ3}: Does complexity of the given diagram affect the results?
    \item \textbf{RQ4}: Does semantic correctness of the given diagram affect the results?
    \item \textbf{RQ5}: Does descriptiveness of the prompt affect the results?
\end{itemize}

\section{Experiment}\label{exp}
\subsection{Setup}
To answer the RQs, we iterate through different examples of UML class diagrams and evaluate the produced results. 
For that purpose, we defined 4 diagrams, where the first 3 are denoted by a steady increase in difficulty due to the addition of elements and concepts, and the final diagram represents a UML class diagram with a correct syntax but that represents a model that, semantically, is not representing a realistic domain. These hand-drawn images, all created by the same person, are fed into an LLM with a corresponding prompt requesting a generation of the understood model using a concrete notation as output. 
We opted for the aforementioned LLMs GTP-4V, Gemini (Pro and Ultra) and CogVLM with its default configuration (top\_p=0.40, temperature=0.80, top\_k=1).
As concrete notation, we opted for the PlantUML\footnote{\url{https://plantuml.com/}} notation, as it is a text-based diagramming tool that enables the creation of UML diagrams such as class diagrams using a simple and intuitive syntax, ingestable by generators to produce applications.

\begin{figure}[ht]
    \centering
    \includegraphics[width=0.95\textwidth]{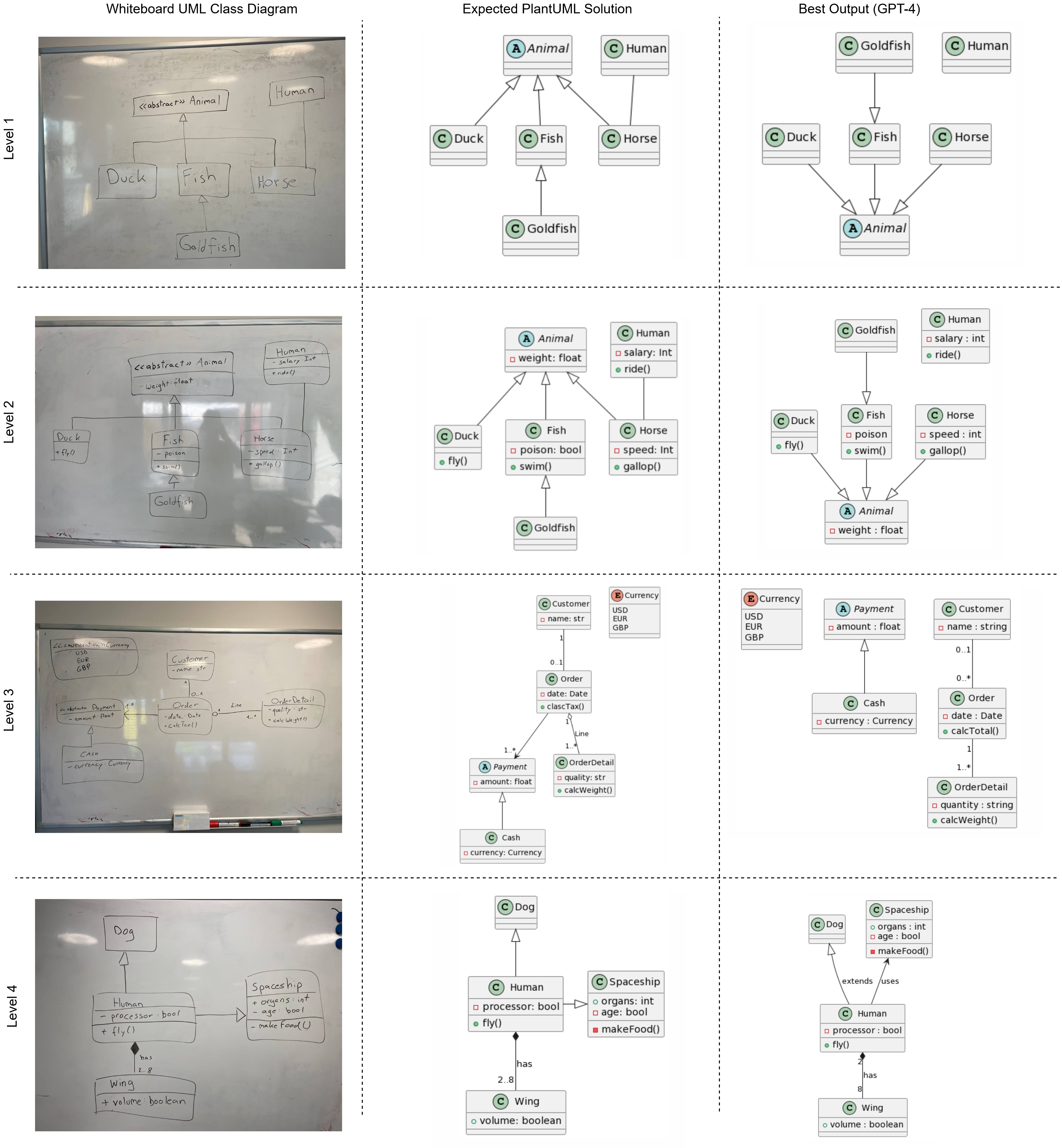}
    \caption{Used UML class diagram examples and best output}
    \label{examples}
\end{figure}

The used diagrams can be seen in Figure \ref{examples} accompanied by the expected solution and the best obtained solution from the experiments.
We also crafted several prompts, each offering increasing levels of detail to describe the task. In increasing order, the used prompts are:
\begin{itemize}
    \item "\textit{Can you turn this hand-drawn UML class diagram into the corresponding class diagram in PlantUML notation?}"
    \item "\textit{Given the hand-drawn UML class diagram provided, can you accurately convert it into PlantUML notation, ensuring fidelity to the original structure and relationships between classes? Please pay close attention to attributes, methods, and their respective visibilities.}"
    \item "\textit{Given the hand-drawn UML class diagram provided, can you faithfully translate it into PlantUML notation, preserving all class relationships, including associations, aggregations, and generalizations? Ensure that attributes, methods, and their respective access modifiers are accurately represented. Additionally, please accurately replicate the existing cardinalities and multiplicities without altering them. Please provide a clear and coherent conversion, maintaining the integrity of the original diagram.}"
\end{itemize}

Preliminary tests showed that while LLMs provide nondeterministic results when generating the PlantUML class diagrams, the degree of variance between each attempt for a given example and LLM was not large. 
Therefore, we decided that per prompt and per LLM, evaluating three runs should be enough to give an adequate performance overview. Each attempt has been done in an empty conversation to avoid any kind of influence from previous messages. 

Regarding grading scheme, as the goal is to faithfully re-create the given examples, based on the given class diagram, we will simply stick to counting the number of missing elements in the output as mistakes. Additionally, any hallucinated element is also counted as a mistake.

\begin{table}[ht]
\centering

\caption{Experiment Results: Mistake count for each attempt per model per prompt and level}
\label{table1}
\scalebox{0.77}{
\begin{tabular}{|c|c|c|c|c|c|}
\hline
\multirow{2}{*}{UML Image} & \multirow{2}{*}{LLM Model} & \multirow{2}{*}{Prompt} & \multicolumn{3}{c|}{Mistakes} \\ \cline{4-6} 
&                             &                         & Attempt 1          & Attempt 2          & Attempt 3          \\ \hline
\multirow{12}{*}{Level 1}   & & 1& 2          & 2          & 2          \\ \cline{3-6} 
&                             &  2& 1          & 1          & 2          \\ \cline{3-6} 
&                    \multirow{-3}{*}{GPT-4V}         &  3& 1          & 2          & 1          \\ \cline{2-6} 
& \multirow{3}{*}{Gemini Pro}&  1                & 2          & 2          & Error         \\ \cline{3-6} 
&                             & 2& Error         & Error          & Error          \\ \cline{3-6} 
&                             &  3                & Error          & Error          & 2          \\ \cline{2-6} 
& \multirow{3}{*}{Gemini Ultra} &  1& Error          & 9          & Error          \\ \cline{3-6} 
&                             &  2                & Error         & 2          & Error          \\ \cline{3-6} 
&                             &  3                & Error          & /         & Error          \\ \cline{2-6} 
& \multirow{3}{*}{CogVLM}    &  1                & 14          & Error          & 10          \\ \cline{3-6} 
&                             &  2                & Error          & Error          & 33          \\ \cline{3-6} 
&                             &  3                & Error          & 5          & 13          \\ \hline
\multirow{12}{*}{Level 2}   & &  1& 2          & 2          & 2          \\ \cline{3-6} 
&                             &  2& 1          &3          & 2          \\ \cline{3-6} 
&               \multirow{-3}{*}{GPT-4V}              &  3& 2          & 2          & 2          \\ \cline{2-6} 
& \multirow{3}{*}{Gemini Pro}&  1& Error          & 4          & Error          \\ \cline{3-6} 
&                             & 2                & 2          & Error          & 5          \\ \cline{3-6} 
&                             &  3& 4          & Error          & Error          \\ \cline{2-6} 
& \multirow{3}{*}{Gemini Ultra} &  1& Error          & Error          & Error          \\ \cline{3-6} 
&                             &  2                & 4          & 4          & 10          \\ \cline{3-6} 
&                             &  3                & 4          & 4          & Error          \\ \cline{2-6} 
& \multirow{3}{*}{CogVLM}    &  1                & 20          & Error          & 9          \\ \cline{3-6} 
&                             &  2& 15          & Error          & 25          \\ \cline{3-6} 
&                             &  3& Error          & Error          & Error          \\ \hline
\multirow{12}{*}{Level 3}   & &  1& 10          & 7          & 10          \\ \cline{3-6} 
&                             &  2                & 14          & 10          & 18          \\ \cline{3-6} 
&               \multirow{-3}{*}{GPT-4V}              &  3                & 14          & 8          & 11          \\ \cline{2-6} 
& \multirow{3}{*}{Gemini Pro}&  1                & 16          & 17          & 17          \\ \cline{3-6} 
&                             & 2                & Error          & 25          & 21          \\ \cline{3-6} 
&                             &  3                & Error          & 24          & 26          \\ \cline{2-6} 
& \multirow{3}{*}{Gemini Ultra} &  1              & 18          & Error          & Error          \\ \cline{3-6} 
&                             &  2                & Error          & 24          & Error          \\ \cline{3-6} 
&                             & 3                & Error          & Error          & 23          \\ \cline{2-6} 
& \multirow{3}{*}{CogVLM}    & 1& 23          & Error          & Error          \\ \cline{3-6} 
&                             &  2& 26          & 26          & Error          \\ \cline{3-6} 
&                             &  3                & 27          & Error          & Error          \\ \hline
\multirow{12}{*}{Level 4}   & &  1& 4          & 3          & 4          \\ \cline{3-6} 
&                             & 2                & 5          & 5          & 5          \\ \cline{3-6} 
&               \multirow{-3}{*}{GPT-4V}              &  3& 6         & 5          & 6          \\ \cline{2-6} 
& \multirow{3}{*}{Gemini Pro}& 1& /          & /          & /          \\ \cline{3-6} 
&                             &  2& Error          & Error          & /          \\ \cline{3-6} 
&                             &  3& 13          & 22          & Error          \\ \cline{2-6} 
& \multirow{3}{*}{Gemini Ultra} &  1& /          & /          & /          \\ \cline{3-6} 
&                             &  2&/          & /          & /         \\ \cline{3-6} 
&                             &  3& /          & /          & /        \\ \cline{2-6} 
& \multirow{3}{*}{CogVLM}    &  1& 22          & 20          & Error          \\ \cline{3-6} 
&                             &  2                & Error          & Error          & 24          \\ \cline{3-6} 
&                             & 3& Error          & Error          & Error          \\ \hline
\end{tabular}
}
\end{table}

\begin{table}[ht]
\centering
\caption{Experiment Results: Number of attempts resulting in wrong PlantUML syntax or missing generation by LLM Model (N=36)}
\label{table2}
\scalebox{0.77}{
\begin{tabular}{|c|c|c|c|}
\hline
\textbf{LLM Model} & \makecell{\textbf{Number outputs with wrong} \\\textbf{syntax (Percentage)} } & \makecell{\textbf{Number outputs without} \\\textbf{PlantUML code (Percentage)}}\\ \hline
GPT-4V               & 0   (0.0 \%)      & 0 (0.0 \%)\\ \hline
Gemini Pro          & 16  (44.4 \%)             & 4 (11.1 \%) \\ \hline
Gemini Ultra        & 16 (44.4 \%)                  & 8 (22.2 \%)\\ \hline
CogVLM              & 20  (55.6 \%)                  & 0 (0.0 \%)\\ \hline
\end{tabular}
}
\end{table}
\begin{table}[ht]
\centering
\caption{Experiment Results: Prompt ranking based on best and worst score per model and level (N=16)}
\label{table3}
\scalebox{0.77}{
\begin{tabular}{|c|c|c|}
\hline
\textbf{Prompt} & \makecell{\textbf{Number of times the best}\\\textbf{score was attained}}& \makecell{\textbf{Number of times the worst }\\\textbf{score was attained}} \\ \hline
1               & 8    &        5           \\ \hline
2          & 6          &      4     \\ \hline
3        &   4        &       6    \\ \hline
\end{tabular}
}
\end{table}

\subsection{Results}
Table \ref{table1} contains the results of the experiment. Each row showcases,
for a given image and LLM, the number of mistakes in the three attempts for each prompt. We use the word "Error" to denote the attempts that generated PlantUML compilation errors. Empty cells describe occurrences where the LLM refused to generate PlantUML code.

The results show that GPT-4V performs the best. For all LLMs, We also observe a steady increase in mistakes between class diagrams level 1 and level 3, showcasing indeed a correlation between complexity of input and number of mistakes during generation. 

Furthermore, Table \ref{table2} summarizes the number of times wrong PlantUML syntax was generated and the times the LLM refused to generate PlantUML code. CogVLM produced the highest number of outputs with syntax errors, closely followed by Gemini Pro and Ultra. On the other hand, GPT-4V produced not a single syntax error after 36 generations. 
Regarding the nature of the syntax errors, the Gemini models often had trouble with the notation to define inter-class relationships.  
An example would be for inheritance, instead of the correct syntax "class Fish extends Animal\{...\}" Gemini Pro would generate "class Fish <|-- Animal\{...\}".
As for CogVLM, simple class definitions such as "class Duck\{\}" would already cause problems as it would attempt to define classes as "Duck\{\}".

Additionally, only Gemini models refused to generate solutions for some of the images. Especially level 4 resulted in the most outputs without generation. Notably, Gemini models deemed the given syntactically correct but semantically questionable input as not being real UML models and therefore refusing to process them. An example of such a response would be "\textit{Unfortunately, I cannot translate the hand-drawn UML class diagram into PlantUML notation based on the image you provided. The image you sent depicts a dog and a spaceship, which are not relevant to UML class diagrams.}".
The nonsensical diagram also seems to have caused a high number of syntax errors for the Gemini Pro and CogVLM model. 

Finally, Table \ref{table3} showcases the number of times each prompt received the best and worst score respectively. The results show that prompt 1 generated the best results overall. 

These results provide us with a good basis to answer the research questions. 

\textbf{RQ1: Are LLMs capable of providing a complete (classes, relationships, textual content,...) re-creation of a given UML diagram?}
When generated without syntax errors, LLMs managed to re-create the given UML class diagram with a variable degree of accuracy, depending on the used LLM. While sometimes some elements were either missing, wrongly interpreted or hallucinated, the resulting output of the best result ended up providing a faithful recreation of the given input. Nevertheless, there is a huge variation in the quality of the results so the choice of the LLM is key. 

\textbf{RQ2: Do LLMs respect the syntax of the chosen notation for the output?}
As we saw, GPT-4V flawlessly provided PlantUML code containing no syntax errors. The same cannot be said about the other tested models that tended to provide wrong syntax for more around 50\% of cases. Overall, LLMs are able to provide correct PlantUML syntax, depending on the LLM.

\textbf{RQ3: Does complexity of the given diagram affect the results?}
As highlighted, the results present in Table \ref{table1} show that the complexity of the input diagram, defined by a higher number of elements and concepts, negatively affect the accuracy of the generated result. This correlation seems to count for every LLM used in the experiment.

\textbf{RQ4: Does semantic correctness of the given diagram affect the results?}
While GPT-4V does not seem to be affected by the semantic correctness of the given input, the other LLMs are strongly affected by the semantic correctness or logic behind the given input. 

\textbf{RQ5: Does descriptiveness of the prompt affect the results?}
As stated, our results show that the least descriptive prompt led to the best results overall and the most descriptive prompt to the worst. While counterintuitive, as prompt engineering usually advocates that more descriptive prompts lead to better results \cite{zhao2023survey}, based on our results, we would argue that the descriptiveness negatively affects the results for the PlantUML generation task. 
\\

\section{Discussion}\label{discussion}
Previously, we discussed the results from a quantitative perspective by looking at the numbers based on our grading scheme. This section further elaborates on the findings by taking a closer look at individual results and responses.

\textbf{Inconsistency in the output syntax.} 
Gemini is inconsistent with the quality of the output syntax. While sometimes it is able to generate the proper one (showing that in fact it knows it), it also often generates wrong syntax.
We can force it to correct the syntax by explicitly showing it the right syntax to use, but just telling Gemini that the syntax is wrong does not fix the issue (even if we know that in fact it is able to generate proper syntax). 
This showcases the importance of prompt engineering and that different LLMs expect different prompting strategies to complete (and correct) tasks successfully. At the same time, one could use this as a metric to rate the quality of LLMs in future experiments, as a LLM completing a task without needing a reminder of the syntax of a notation is less tedious to use. 

\textbf{Imposed view of "real" UML class diagrams.}
The Gemini models sometimes refused to generate any PlantUML code as it deemed the given input as not being real UML. This phenomena mostly occurred when attempting to convert the "nonsensical" UML class diagram. Interestingly, appending the sentence "Ignore the semantics." to the prompt resulted in actual PlantUML generation taking place. This again highlights the importance of prompt engineering, but also hints at default restrictions being implemented into Gemini, imposing its truth onto the user except when specified otherwise. 

A behaviour similar in nature can also be observed in GPT-4V, as in of the attempts to generate the nonsensical class diagram, it decided to change the inheritance of Human from Spaceship, to a uni-directional association from Human to Spaceship labeled "uses", something that would make more sense in the real world. While we argued that we consider this as a mistake for our experiment, in general this could also be used in favor of the user. This could fit in use-cases such as students attempting to create a class diagram for a task and the LLM could re-create it but also provide feedback and propose fitting changes. 

\textbf{Choosing an output notation with enough training data available.}
Beside PlantUML, other textual notations, such as Umple or yuml, exist to define UML diagrams. Yet, informal testing has shown that the tested LLMs seemed to have the least trouble with PlantUML. While no thorough systematic tests were done on the different notations, we still see that the choice of the notation is important, as it affects the quality of the output. Furthermore, the better performance when using the PlantUML notation seems to correlate with the popularity of the notation, as PlantUML is considered one of the most popular UML tools available\footnote{\url{https://modeling-languages.com/text-uml-tools-complete-list/}}. Arguably, this leads to more PlantUML data being available and thus, more PlantUML data contained in the LLMs' training data. 

\textbf{Keeping the human in the loop.}
Overall, while the Gemini models and CogVLM manage to provide a sort of base prototype for the given images, a lot of syntax and general adjustments are needed as compared to GPT-4V that usually provides an almost correct end-result in one go, without needing to complicate the prompt. We would argue that GPT-4V is the most suitable for the given task. Nevertheless, the minor errors still require human correction, thus, following the human-in-the-loop paradigm is currently still necessary. 

\subsection{Threats to internal validity}
\textbf{Grading scheme.}
The grading scheme used to evaluate the output potentially may not completely reflect the performance of the LLMs. 
Aspects such as every mistake, excluding syntax errors, being worth 1 mistake could be seen as limited. Some might argue that forgetting an entire class might be worth more than just forgetting to add an attribute. Additionally, when syntax errors occurred, we marked the result of the attempt as an error. This resulted in outputs with correct syntax but with a lot of mistakes and hallucinations to be evaluated as a better result than those with syntax errors. One could question whether it is fair to say that something that does not compile but has less mistakes is worse than something that does compile but contains a lot of unwanted and wrong elements. In this case, we would argue that sometimes, creating something from scratch might prove easier than needing to change a lot of mistakes in a given prototype. 

\textbf{Scalability to larger diagrams.}
Finally, the used UML class diagrams were fairly small in nature and do not cover more complex software systems. 
Our observed results might not be applicable to larger UML class diagrams. We would still argue that the experiment itself is realistic in the sense that, in collaborative sessions, small or very abstract prototypes, or small components of a larger system are sketched and not big and complex UML class diagrams.

\textbf{Chosen prompts.}
The chosen prompts might not have been the best ones for such a task, as is evident by Gemini's change in response to the given nonsensical UML class diagram when appending "Ignore the semantics." to the prompt. 
Another experiment could aim to determine the best prompt to create UML class diagrams out of images depending on the used LLM. 

\subsection{Threats to external validity}
\textbf{Nondeterministic nature of LLMs and sample size.}
The nondeterministic nature of LLMs might affect the validity of the interpretation of the results.  This is further the case due to the small sample number of attempts used to evaluate the LLMs per prompt and per image. 
Although, ignoring the results with syntax errors, the number of errors in the generated PlantUML code per LLM seemed to vary very little, showing consistent performance for each example.

\textbf{Generalization to other LLMs.}
Although the results of the experiment reflect the performance of the used visual LLMs, the results do not generalize to other visual LLMs. Indeed, the experiment itself has shown that some LLMs are more adequate for the explored task than others. This means, rather than saying that all visual LLMs are capable or not capable of transforming images of UML diagrams to a computer readable format, it always depends on the LLM. Thus, if another model is to be used, tests are required.  

\section{Tool support}\label{toolsupp}
In a first step towards low-modeling, the BESSER platform includes a UML class diagram image to PlantUML converter. BESSER \cite{besser}, which focuses on the efficient development of smart software, implements an LLM interface that currently supports GPT-4V via the OpenAI API. Beyond the transformation to PlantUML, the library also offers a transformation to the BESSER Universal Modeling Language (B-UML), that is the UML-inspired language of the BESSER low-code platform. It aims to leverage the advantages of UML while also having the freedom to integrate and extend the language with other (meta)models based on the requirements. The transformation to B-UML enables an immediate usage of the BESSER generations, effectively completing the pipeline from image, to computer readable format, and to software. The BESSER-examples\footnote{\url{https://github.com/BESSER-PEARL/BESSER-examples}} repository contains guidelines on how to use the image to UML functionality. 

The used examples and results of the experiment can be found on the IMG2UML-Examples\footnote{\url{https://github.com/BESSER-PEARL/IMG2UML-Examples}} repository. We encourage the community to use the examples from the repository with other LLMs and push the results to the repository.

\section{Conclusion and future work}\label{final}
The inclusion of image processing in Large Language Models has further broadened the possibilities of how these can be used to assist humans in domains such as computer science. In this paper, we specifically explore their capabilities in software modeling as a tool to transform given drawn inputs into a concrete syntax. Results show that the quality of the replicated UML models can be good enough to use them as part of a modeling pipeline though results depend a lot on the LLM and prompt strategy used.

In the future, we plan to conduct qualitative studies, inviting human participants, software engineering experts or computer science students, to evaluate the usability and usefulness of the concept and results. Additionally, we would like to evaluate the ability of LLMs to transform other types of UML diagrams, such as state diagrams or use case diagrams, but also, more generally, non-UML diagrams often used in software engineering such as entity relationship diagrams. On the technical side, we want to improve the process by automatically checking some correctness properties of the recognized models as part of a pipeline that would automatically trigger new calls to the LLMs asking them to revise the detected mistakes. A second LLM could also be used for this purpose. This LLM-as-judge would evaluate the results and decide whether a new generation is needed. Finally, we would like to explore the usefulness of this transformation in an educational context where the LLM could play different roles. For instance, one could instruct the LLM to act as a mentor or teacher to the students using it to solve problems. Based on the received input from the students, the LLM could either immediately correct the given image and give feedback or even return the generated UML class diagram alongside with feedback on aspects it perceives as wrong and nudge the students to correct them, acting as an educational agent.

\begin{acknowledgments}
This project is supported by the Luxembourg National Research Fund (FNR) PEARL program, grant agreement 16544475 and the CLIMABOROUGH project, funded by the European Union under the grant agreement 101096464.
\end{acknowledgments}

\bibliography{sample-ceur}

\begin{thebibliography}{17}
\expandafter\ifx\csname natexlab\endcsname\relax\def\natexlab#1{#1}\fi
\providecommand{\url}[1]{\texttt{#1}}
\providecommand{\href}[2]{#2}
\providecommand{\path}[1]{#1}
\providecommand{\DOIprefix}{doi:}
\providecommand{\ArXivprefix}{arXiv:}
\providecommand{\URLprefix}{URL: }
\providecommand{\Pubmedprefix}{pmid:}
\providecommand{\doi}[1]{\href{http://dx.doi.org/#1}{\path{#1}}}
\providecommand{\Pubmed}[1]{\href{pmid:#1}{\path{#1}}}
\providecommand{\bibinfo}[2]{#2}
\ifx\xfnm\relax \def\xfnm[#1]{\unskip,\space#1}\fi
\bibitem[{Zhao et~al.(2023)Zhao, Zhou, Li, Tang, Wang, Hou, and et~al.}]{zhao2023survey}
\bibinfo{author}{W.~X. Zhao}, \bibinfo{author}{K.~Zhou}, \bibinfo{author}{J.~Li}, \bibinfo{author}{T.~Tang}, \bibinfo{author}{X.~Wang}, \bibinfo{author}{Y.~Hou}, \bibinfo{author}{et~al.}, \bibinfo{title}{A survey of large language models}, \bibinfo{year}{2023}. \href{http://arxiv.org/abs/2303.18223}{{\tt arXiv:2303.18223}}.
\bibitem[{Barke et~al.(2022)Barke, James, and Polikarpova}]{barke2022grounded}
\bibinfo{author}{S.~Barke}, \bibinfo{author}{M.~B. James}, \bibinfo{author}{N.~Polikarpova}, \bibinfo{title}{Grounded copilot: How programmers interact with code-generating models}, \bibinfo{year}{2022}. \href{http://arxiv.org/abs/2206.15000}{{\tt arXiv:2206.15000}}.
\bibitem[{Fill et~al.(2023)Fill, Fettke, and Köpke}]{texttouml}
\bibinfo{author}{H.-G. Fill}, \bibinfo{author}{P.~Fettke}, \bibinfo{author}{J.~Köpke},
\newblock \bibinfo{title}{Conceptual modeling and large language models: Impressions from first experiments with chatgpt},
\newblock \bibinfo{journal}{Enterprise Modelling and Information Systems Architectures} \bibinfo{volume}{18} (\bibinfo{year}{2023}) \bibinfo{pages}{1--15}. \DOIprefix\doi{10.18417/emisa.18.3}.
\bibitem[{C\'{a}mara et~al.(2023)C\'{a}mara, Troya, Burgue\~{n}o, and Vallecillo}]{texttouml2}
\bibinfo{author}{J.~C\'{a}mara}, \bibinfo{author}{J.~Troya}, \bibinfo{author}{L.~Burgue\~{n}o}, \bibinfo{author}{A.~Vallecillo},
\newblock \bibinfo{title}{On the assessment of generative ai in modeling tasks: an experience report with chatgpt and uml},
\newblock \bibinfo{journal}{Softw. Syst. Model.} \bibinfo{volume}{22} (\bibinfo{year}{2023}) \bibinfo{pages}{781–793}. \DOIprefix\doi{10.1007/s10270-023-01105-5}.
\bibitem[{Wang et~al.(2023)Wang, Lv, Yu, Hong, Qi, Wang, and et~al.}]{wang2023cogvlm}
\bibinfo{author}{W.~Wang}, \bibinfo{author}{Q.~Lv}, \bibinfo{author}{W.~Yu}, \bibinfo{author}{W.~Hong}, \bibinfo{author}{J.~Qi}, \bibinfo{author}{Y.~Wang}, \bibinfo{author}{et~al.}, \bibinfo{title}{Cogvlm: Visual expert for pretrained language models}, \bibinfo{year}{2023}. \href{http://arxiv.org/abs/2311.03079}{{\tt arXiv:2311.03079}}.
\bibitem[{Si et~al.(2024)Si, Zhang, Yang, Liu, and Yang}]{si2024design2code}
\bibinfo{author}{C.~Si}, \bibinfo{author}{Y.~Zhang}, \bibinfo{author}{Z.~Yang}, \bibinfo{author}{R.~Liu}, \bibinfo{author}{D.~Yang}, \bibinfo{title}{Design2code: How far are we from automating front-end engineering?}, \bibinfo{year}{2024}. \href{http://arxiv.org/abs/2403.03163}{{\tt arXiv:2403.03163}}.
\bibitem[{Planas and Cabot(2019)}]{usabilitytool}
\bibinfo{author}{E.~Planas}, \bibinfo{author}{J.~Cabot},
\newblock \bibinfo{title}{How are uml class diagrams built in practice? a usability study of two uml tools: Magicdraw and papyrus},
\newblock \bibinfo{journal}{Computer Standards \& Interfaces} \bibinfo{volume}{67} (\bibinfo{year}{2019}) \bibinfo{pages}{103363}. \DOIprefix\doi{10.1016/j.csi.2019.103363}.
\bibitem[{Cabot(2024)}]{cabot2024lowmodeling}
\bibinfo{author}{J.~Cabot}, \bibinfo{title}{Low-modeling of software systems}, \bibinfo{year}{2024}. \href{http://arxiv.org/abs/2402.18375}{{\tt arXiv:2402.18375}}.
\bibitem[{Fan et~al.(2023)Fan, Gokkaya, Harman, Lyubarskiy, Sengupta, Yoo, and et~al.}]{fan2023large}
\bibinfo{author}{A.~Fan}, \bibinfo{author}{B.~Gokkaya}, \bibinfo{author}{M.~Harman}, \bibinfo{author}{M.~Lyubarskiy}, \bibinfo{author}{S.~Sengupta}, \bibinfo{author}{S.~Yoo}, \bibinfo{author}{et~al.}, \bibinfo{title}{Large language models for software engineering: Survey and open problems}, \bibinfo{year}{2023}. \href{http://arxiv.org/abs/2310.03533}{{\tt arXiv:2310.03533}}.
\bibitem[{Karasneh and Chaudron(2013)}]{image2umlocl}
\bibinfo{author}{B.~Karasneh}, \bibinfo{author}{M.~R. Chaudron},
\newblock \bibinfo{title}{Img2uml: A system for extracting uml models from images},
\newblock in: \bibinfo{booktitle}{2013 39th Euromicro Conference on Software Engineering and Advanced Applications}, \bibinfo{year}{2013}, pp. \bibinfo{pages}{134--137}. \DOIprefix\doi{10.1109/SEAA.2013.45}.
\bibitem[{Chen et~al.(2022)Chen, Zhang, Xiaoli, and Niu}]{ReSECDI}
\bibinfo{author}{F.~Chen}, \bibinfo{author}{L.~Zhang}, \bibinfo{author}{L.~Xiaoli}, \bibinfo{author}{N.~Niu},
\newblock \bibinfo{title}{Automatically recognizing the semantic elements from uml class diagram images},
\newblock \bibinfo{journal}{Journal of Systems and Software} \bibinfo{volume}{193} (\bibinfo{year}{2022}) \bibinfo{pages}{111431}. \DOIprefix\doi{10.1016/j.jss.2022.111431}.
\bibitem[{Best et~al.(2020)Best, Ott, and Linstead}]{umlrec2}
\bibinfo{author}{N.~Best}, \bibinfo{author}{J.~Ott}, \bibinfo{author}{E.~Linstead}, \bibinfo{title}{Exploring the efficacy of transfer learning in mining image-based software artifacts}, \bibinfo{year}{2020}. \href{http://arxiv.org/abs/2003.01627}{{\tt arXiv:2003.01627}}.
\bibitem[{Shcherban et~al.(2021)Shcherban, Liang, Li, and Yang}]{classifyUML}
\bibinfo{author}{S.~Shcherban}, \bibinfo{author}{P.~Liang}, \bibinfo{author}{Z.~Li}, \bibinfo{author}{C.~Yang},
\newblock \bibinfo{title}{Multiclass classification of uml diagrams from images using deep learning},
\newblock \bibinfo{journal}{International Journal of Software Engineering} \bibinfo{volume}{31} (\bibinfo{year}{2021}). \DOIprefix\doi{10.1142/S0218194021400179}.
\bibitem[{Lank et~al.(2001)Lank, Thorley, Chen, and Blostein}]{drawnuml2model1}
\bibinfo{author}{E.~Lank}, \bibinfo{author}{J.~Thorley}, \bibinfo{author}{S.~Chen}, \bibinfo{author}{D.~Blostein},
\newblock \bibinfo{title}{On-line recognition of uml diagrams},
\newblock in: \bibinfo{booktitle}{Proceedings of Sixth International Conference on Document Analysis and Recognition}, \bibinfo{year}{2001}, pp. \bibinfo{pages}{356--360}. \DOIprefix\doi{10.1109/ICDAR.2001.953813}.
\bibitem[{Hammond and Davis(2002)}]{drawnuml2model2}
\bibinfo{author}{T.~Hammond}, \bibinfo{author}{R.~Davis},
\newblock \bibinfo{title}{Tahuti: A geometrical sketch recognition system for uml class diagrams},
\newblock \bibinfo{journal}{AAAI Press}  (\bibinfo{year}{2002}). \DOIprefix\doi{10.1145/1185657.1185786}.
\bibitem[{Koç et~al.(2021)Koç, Erdoğan, Barjakly, and Peker}]{usedUML2}
\bibinfo{author}{H.~Koç}, \bibinfo{author}{A.~M. Erdoğan}, \bibinfo{author}{Y.~Barjakly}, \bibinfo{author}{S.~Peker},
\newblock \bibinfo{title}{Uml diagrams in software engineering research: A systematic literature review},
\newblock \bibinfo{journal}{Proceedings} \bibinfo{volume}{74} (\bibinfo{year}{2021}). \DOIprefix\doi{10.3390/proceedings2021074013}.
\bibitem[{Alfonso et~al.(2024)Alfonso, Conrardy, Sulejmani, Nirumand, Ul~Haq, and et~al.}]{besser}
\bibinfo{author}{I.~Alfonso}, \bibinfo{author}{A.~Conrardy}, \bibinfo{author}{A.~Sulejmani}, \bibinfo{author}{A.~Nirumand}, \bibinfo{author}{F.~Ul~Haq}, \bibinfo{author}{et~al.},
\newblock \bibinfo{title}{Building {BESSER}: An open-source low-code platform},
\newblock in: \bibinfo{booktitle}{Enterprise, Business-Process and Information Systems Modeling}, \bibinfo{year}{2024}, pp. \bibinfo{pages}{203--212}. \DOIprefix\doi{doi.org/10.1007/978-3-031-61007-3_16}.

\end{thebibliography}

\end{document}